\documentclass[twocolumn,showpacs,preprintnumbers,amsmath,amssymb]{revtex4}
\usepackage{graphicx}% Include figure files
\usepackage{dcolumn}% Align table columns on decimal point
\usepackage{bm}% bold math

\begin{document}
    
% old title :
% \title{Origins of memory in an aging molecular glass}

\title{Memory in an aging molecular glass}
\author{H. Yardimci and R. L. Leheny}
\affiliation{Department of Physics and Astronomy, Johns Hopkins
University,
Baltimore, MD 21218, USA}
\date{\today}

\begin{abstract} 
    
    The dielectric susceptibility of the
    molecular liquid sorbitol below its calorimetric glass transition 
    displays memory strikingly 
    similar to that of a variety of glassy materials.  During a temporary 
    stop in cooling, the susceptibility changes with time, and upon reheating the 
    susceptibility retraces these changes.  To investigate the 
    out-of-equilibrium state of the liquid as it displays this memory, 
    the heating stage of this cycle is interrupted and the 
    subsequent aging characterized.  At temperatures above that of the original
    cooling stop, the liquid enters a state on heating with an effective age 
    that is proportional to the duration of the stop, while at lower 
    temperatures no effective age can be assigned
    and subtler behavior emerges.  These results, which reveal 
    differences with memory displayed by spin glasses, are 
    discussed in the context of the liquid's energy landscape.

\end{abstract}


\maketitle

For many condensed matter systems with disorder, a failure to achieve equilibrium 
can dominate behavior, and understanding the out-of-equilibrium states becomes 
crucial to their accurate description.  During the protracted equilibration 
of such systems, dynamic correlation functions and
response functions are dependent on time and 
thermal history.  This behavior, known as aging, appears in a broad variety 
of disordered materials including polymers~\cite{struik,matsuoka},  
spin glasses~\cite{vincent-review,nordblad-review}, molecular glasses~\cite{leheny},
gels~\cite{cipelletti,harden} and disordered 
ferroelectrics~\cite{ferroelectric-aging1,ferroelectric-aging2}.  The widespread 
occurrence of aging has led to theoretical pictures that explain 
the phenomenon as a general consequence of disorder~\cite{bouchaud} and has further 
motivated work into how one might generalize concepts of equilibrium 
statistical mechanics, such as the fluctuation-dissipation theorem, to 
the aging ``state''~\cite{cugliandolo,grigera}. In addition, specific 
features of aging, particularly in spin glasses, have been 
interpreted as providing insight into the underlying equilibrium 
physics of disordered systems.   

An intriguing property of many aging systems is the memory of their thermal 
histories
that their bulk thermodynamic properties and response functions display. 
In this paper, we describe dielectric susceptibility experiments 
that probe memory in the molecular liquid sorbitol at temperatures below 
its calorimetric glass transition.  These measurements follow an experimental 
procedure, originally developed for spin glasses~\cite{jonason},
that dramatically illustrates memory in aging systems.  
Consistent with studies on other glassy materials, specifically 
a polymer~\cite{bellon} and an orientational glass~\cite{doussineau}, 
we find that the memory 
displayed by glassy sorbitol appears strikingly similar to that of 
spin glasses.  Such apparently universal behavior suggests generic underlying 
mechanisms governing memory, with important implications for 
understanding aging as a whole.  Therefore, to obtain a more 
comprehensive perspective of sorbitol 
as it displays memory, we have performed additional measurements in which we 
interrupt the memory experiments to track the aging of the system.  
The observed aging provides a characterization of the 
out-of-equilibrium states that the glass enters during the memory 
experiment.  As explained below, these measurements elucidate the significance of
memory in sorbitol and help to identify  
important differences between the out-of-equilibrium states of 
structural glasses and spin glasses.

%The susceptibility, 
%$\epsilon$($\nu$)=$\epsilon$'($\nu$)+$i\epsilon$''($\nu$), of sorbitol was 
%determined by measuring the complex impedance of a capacitor filled with the 
%liquid relative to impedance of empty capacitor. The capacitor 
%consisted of two aluminum electrodes having a seperation of 
%? $\mu$m, as shown in the inset of Fig. 1(a). Crystalline sorbitol (99+ \%), obtained from Aldrich Chemical Co., 
%was heated well above its melting temperature in the cup-shaped bottom 
%elctrode of the capacitor until the sample was thoroughly melted. The 
%top electrode was then pressed into the molten sorbitol and the 
%sample was further heated above the melting point for several hours. The sample 
%was then rapidly been cooled to obtain glassy sorbitol.

%The capacitor was placed within a heating coil in 
%a sealed copper can with dry nitrogen environment, and the can was
%submerged in a liquid nitrogen filled dewar. 
%The resistance through heating coil was controlled by a PID 
%temperature controller, and the temperature of the capacitor was 
%controlled with an accuracy 
%of $\pm$0.02 K. 

%Low frequency susceptibility measurements were performed using a 
%Stanford 850 digital lock-in amplifier in conjuction with a Keithley 
%428 current amplifier. The output of the lock-in amplifier was placed 
%accross the capacitor in series with the current amplifier. We 
%obtained the impedance of the circuit by measuring the output of 
%current amplifier with the lock-in. Above 10$^4$ Hz, impedance 
%measurements were made using a Hewlett Packard 4284A LCR meter.

Following the protocol established in Ref.~\cite{jonason}, 
we begin each 
measurement with sorbitol at $T=276$ K, well above 
the calorimetric glass transition, $T_{g}=268$ K.  To create a 
reference measurement, we cool the liquid
at a fixed rate of 0.13 K/minute to 160 K  and then heat back to 276 K at the 
same rate.  Figure 1(a) shows the resulting values of the imaginary 
part of the dielectric susceptibility,
$\epsilon$''($\nu$), at $\nu$=0.79 Hz as a function of temperature. The 
hysteresis between cooling and reheating indicates that the system ages
during the time spent below $T_{g}$.  As previous studies of supercooled 
liquids have demonstrated, $\epsilon$''($\nu$) at fixed $\nu$ 
(above the peak frequency for the alpha relaxation) 
decreases with time during aging~\cite{leheny}, consistent with the 
hysteresis in Fig.~1(a).  The broad peak in $\epsilon$''(0.79 Hz) centered near 190 
K is a Johari-Goldstein beta relaxation.  The aging of this 
beta peak tracks that of the primary alpha relaxation; details of this
behavior will be provided elsewhere~\cite{beta}. 

To characterize the memory, the cooling and reheating is 
repeated except that the cooling is temporarily stopped at
an intermediate temperature, $T_{stop}$.  Fig.~1(b) displays 
values of $\epsilon$''(0.79 Hz) versus temperature 
for a measurement that includes a temporary stop in the 
cooling for $t_{stop}=12$ hours at $T_{stop}=230$ K.  
The solid lines in the figure are the results without 
the stop taken from Fig.~1(a).  To highlight the effect of a 
stop, the ratio of the susceptibility
measured on cooling with a stop, denoted $\epsilon$''$_{stop}$, to 
that without a stop,  denoted $\epsilon$''$_{ref}$, is plotted in Fig.~2(a).  
Susceptibility ratios from measurements performed with stops at
$T_{stop}= 230$ K, 215 K, and 200 K for 12 hours are plotted.  For $T>T_{stop}$,
$\epsilon$''$_{stop}$/$\epsilon$''$_{ref} = 1$, as expected. 
The drop in the ratio at $T_{stop}$ represents the aging 
that occurs during the stop.  Upon further cooling $\epsilon$''$_{stop}$/$\epsilon$''$_{ref}$ 
returns to a value near one, suggesting that the out-of-equilibrium state becomes 
decreasingly affected by the aging during the stop as the temperature 
deviates further from $T_{stop}$; that is, suggesting that the liquid ``rejuvenates''. 
Fig.~2(b) shows the ratio of susceptibilities on reheating with and without the previous
stops on cooling.
Much like the cooling ratio, the heating ratio goes through a minimum at a temperature very 
close to $T_{stop}$ before returning to one at higher 
temperatures. This clear duplication, or memory, of the minimum near $T_{stop}$
suggests that the out-of equilibrium state of the liquid at temperatures 
in the vicinity of $T_{stop}$ remains 
sensitive to the previous aging, 
while at temperatures away from 
$T_{stop}$, the state remains indifferent to the aging.  
Longer waiting times at $T_{stop}$ increase the depth 
of the minimum, leaving its position in temperature unchanged.  

\begin{figure}
    \centering\includegraphics[scale=1.0]{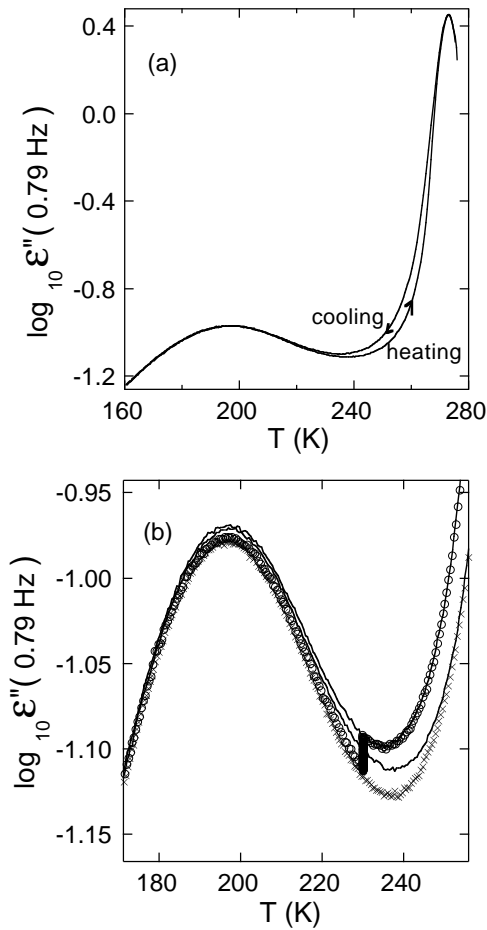}
\caption{\label{fig1} (a) The imaginary part of the dielectric susceptibility,
$\epsilon$''($\nu$), at $\nu$=0.79 Hz for sorbitol ($T_{g}=268$ K) 
as a function of temperature during cooling and reheating at a 
rate of 0.13 K/minute.
(b)  The susceptibility during cooling (circles) and reheating (X's) 
on an expanded scale for the case in which 
the cooling is stopped at $T_{stop}=$230 K for $t_{stop}=$ 12 hours.  
The solid lines show the results without any stop.}
\end{figure}

As mentioned above, these results for sorbitol, a molecular liquid, 
appear strikingly similar to those observed in other aging 
systems~\cite{jonason,bellon,doussineau}, suggesting common underlying mechanisms.  
A central feature of all these memory studies is that the values of 
susceptibility at temperatures away from $T_{stop}$ rapidly approach 
each other regardless of the aging that occurs during the stop.
However, as first demonstrated by Kovacs~\cite{kovacs}, measured 
values of thermodynamic 
quantities for aging materials with different thermal histories
can coincide even while they are in very different out-of-equilibrium states.
Therefore, to investigate the significance of the memory effect in 
sorbitol, we have determined how the liquid ages when the subsequent reheating is interrupted 
following the cooling stop.  These aging measurements provide
a method to characterize the out-of-equilibrium state of the liquid as
it retraces the ratio minimum in Fig.~2(b).  Significantly, we find that the behavior 
observed in these measurements is sensitive to whether the liquid is 
reheated to a temperature that is greater or less than $T_{stop}$. 

\begin{figure}
    \centering\includegraphics[scale=1.0]{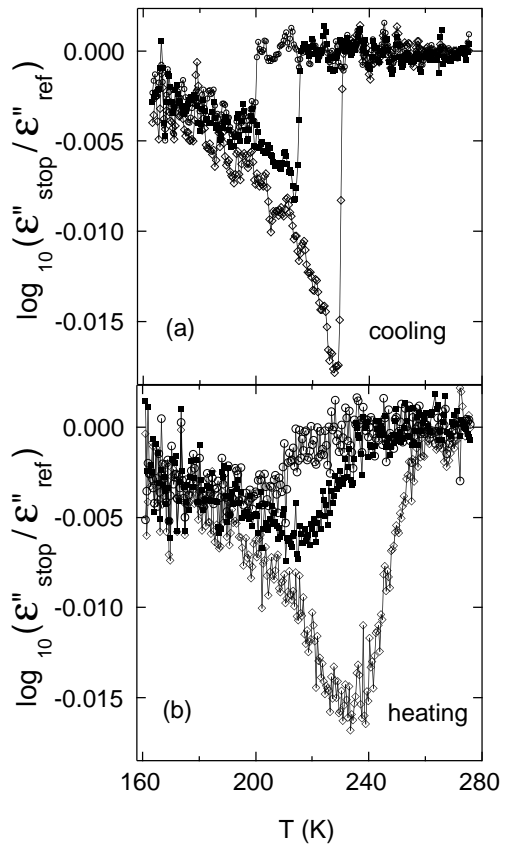}
\caption{\label{fig2} (a) The ratio of $\epsilon$''(0.79 Hz) measured on cooling with 
a stop, $\epsilon$''$_{stop}$, to its value measured without of stop, 
$\epsilon$''$_{ref}$, for stops of duration 
$t_{stop}=$ 12 hours at $T_{stop}=$ 200 K (open circles), 215 K 
(solid squares) and 230 K (open diamonds).  (b) The ratio of the
susceptibilities measured on 
reheating.  The ratio on reheating displays a minimum near the 
temperature of the cooling stop.}
\end{figure}

Figure 3(a) shows a representative set of results for the case when 
the heating is interrupted at a temperature $T>T_{stop}$.  The figure displays
$\epsilon$''(0.79 Hz) as a function 
of aging time, $t_{a}$, after heating to $T=220$ K following cooling 
stops at $T_{stop}=215$ K of various durations.  The
behavior of the susceptibility after the heating is interrupted, 
$\epsilon$''(0.79 Hz, $t_{a}$), depends on
the duration of the cooling stop, $t_{stop}$.
However, as illustrated in Fig.~3(b), $\epsilon$''(0.79 Hz, $t_{a}$)
corresponding to different values of $t_{stop}$ can be superposed
through an additive shift of the aging time.  The values of 
these shifts, $t_{shift}$, are simply
proportional to $t_{stop}$, with a proportionality constant that 
depends on temperature.  For example, for the results in Fig.~3(b)
corresponding to $T=220$ K and $T_{stop}=215$ K, the aging curves 
are collapsed
using values of $t_{shift}$ given by 
$t_{shift}/t_{stop}=0.53$.  Thus, upon reheating during the memory 
experiments to temperatures $T \geq T_{stop}$, sorbitol enters the 
same out-of-equilibrium state that it would have reached with no cooling stop 
but with aging at $T$ for $t_{shift}$.  In other words, the liquid 
after reheating to $T \geq T_{stop}$
has its effective age increased by $t_{shift}$ due to the previous 
cooling stop.

\begin{figure}
    \centering\includegraphics[scale=1.0]{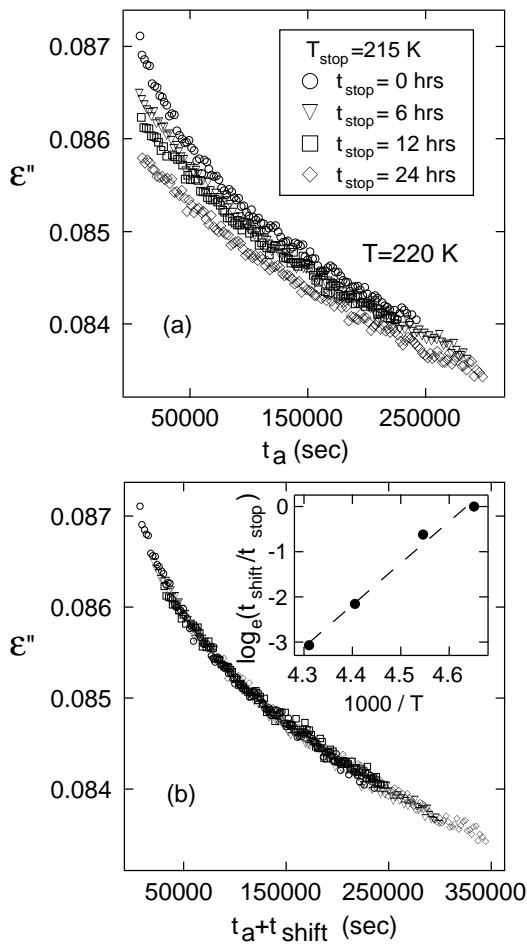}
\caption{\label{fig3} (a) $\epsilon$''(0.79 Hz) versus time, $t_{a}$, following an
interruption in reheating at 220 K during memory experiments that include 
a cooling stop at $T_{stop}=$ 215 K of duration $t_{stop}=$ 0 hours (circles), 6 
hours (triangles), 12 hours (squares), and 24 hours (diamonds).  (b) 
$\epsilon$''(0.79 Hz,$t_{a}$) for these measurements
with the time axis shifted by $t_{shift}$ to collapse the results onto a single curve.  
The inset shows the ratio 
$t_{shift}/t_{stop}$ leading to such collapse of aging curves
as a function of inverse temperature for $T>T_{stop}$, where $T_{stop}=$215 K.}
\end{figure}

The inset of Fig.~3(b) shows the ratios $t_{shift}/t_{stop}$ used to 
collapse aging curves corresponding to different $t_{stop}$ 
as a function of temperature
for $T \geq T_{stop}$, where $T_{stop}=215$ K.  
The temperature dependence 
displayed by $t_{shift}/t_{stop}$ is fully consistent with 
aging as a thermally activated process within a fixed (i.e., 
temperature independent) energy landscape.  Within such a picture 
$t_{shift}/t_{stop}$ is expected to vary with temperature as

\begin{equation}
  log_{e}[t_{shift}/t_{stop}] = (T_{stop}/T - 1)log_{e}[t_{stop}/\tau_{0}]
\end{equation} 

\noindent
where $\tau_{0}$ is a microscopic time scale~\cite{vincent-review,refregier}.
The results of a fit to Eq.~(1), shown with the dashed line in the inset, 
give $\tau_{0} \sim 10^{-14 \pm 1}$ s, a physically 
plausible time scale.  Expressed another way, for $t_{stop} = $ 86400 
s and $\tau_{0}$ constrained to the realistic value 
$\tau_{0} = 10^{-13}$ s, Eq.~(1) predicts 
$t_{shift}$ = 34000 s for T = 220 K, 9700 s for T = 227 K, and 4200 s 
for T = 232 K, which compare closely with the measured values of 46000 s, 10000 
s and 4000 s, respectively.   
%
%For comparison, the 
%Johari-Goldstein beta peak in sorbitol displays an Arrhenius temperature 
%dependence, $log_{e}(\nu_{JG}) \sim -U_{JG}/T$, 
%with an activation energy $U_{JG} = 7400 \pm 600$ K~\cite{beta,wagner}.  
%The similarity between these activation 
%energies suggests a link between aging dynamics and the beta relaxation 
%in this structural glass.  Similar studies of other glass-forming 
%liquids would help to determine whether this approximate agreement is 
%purely coincidental or
%represents a true connection between the mechanisms of beta 
%relaxation and aging.  
We emphasize that this success of activated scaling for a 
molecular glass contrasts strongly with the behavior of Heisenberg spin glasses, 
where the ``superactivated'' nature of the aging is well 
documented~\cite{vincent-review,refregier}.  This differing 
sensitivity to temperature in the aging of spin glasses and structural 
glasses may also 
be reflected in the contrasting widths in temperature over which the 
memory effects occur.  For example, the deviations, 
$\epsilon$''$_{stop}$/$\epsilon$''$_{ref} < 1$,
observed in Fig.~2(b) extend over more 
than 33\% in absolute temperature, a range comparible with that observed in 
ferroelectric~\cite{doussineau} and polymer~\cite{bellon,bellonEPJB} glasses.  
In contrast, the deviations in spin glasses can be much sharper, in 
some cases less than 10\% of absolute temperature ~\cite{jonason}.

%The activated temperature dependence of $t_{shift}/t_{stop}$ for $T>T_{stop}$ further 
%leads directly to the rapid return of $\epsilon''_{stop}$/$\epsilon''_{ref}$ to one at 
%high temperature in the memory experiments.  Due to the success of this scaling, the 
%ratio of susceptibilities on heating can be approximated as 
%
%\begin{equation}
%log_{e}\left[\epsilon''_{stop}/\epsilon''_{ref}\right] \approx 
%t_{shift}\left[\frac{d}{dt_{a}}log_{e}(\epsilon''(t_{a}))\right].   
%\end{equation} 
%
%\noindent
%Measurements, such as those displayed in Fig.~3(b), show that
%$\frac{d}{dt_{a}}log_{e}(e''(t_{a}))$ is a relatively weak function of 
%temperature ($\frac{d}{dt_{a}}log_{e}(e''(t_{a}))\sim 
%exp(-U_{1}/T)$ with $U_{1}=5200 \pm 200$), so that the temperature 
%dependence of $log_{e}\left[\epsilon''_{stop}/\epsilon''_{ref}\right]$ 
% above $T_{stop}$ is dictated largely by $t_{shift}$.

While further measurements indicate that the scaling through $t_{shift}$ 
illustrated in Fig.~3(b) succeeds for all $T \gtrsim T_{stop}$, 
regardless of $T_{stop}$ or $t_{stop}$, it breaks down for heating 
interruptions at temperatures below $T_{stop}$.  
Figure 4(a) shows a representative set of results for 
$\epsilon''$(0.79 Hz, $t_{a}$) when the heating is interrupted at 185 K 
following cooling stops at $T_{stop}=215$ K of various durations.  We 
note that Fig.~4(a) displays significant aging at 185 K despite the 
different curves in Fig.~1(b) having essentially 
converged at this low temperature.  Also included 
in the figure is $\epsilon''$(0.79 Hz, $t_{a}$) at 185 K following a 
12 hour cooling stop at $T_{stop}=260$ K.  The absolute values of $\epsilon''$(0.79 Hz)
on heating to 185 K depend on the specifics of the cooling
stop; however, the shapes of $\epsilon''$(0.79 Hz, $t_{a}$) are independent of 
both $t_{stop}$ and $T_{stop}$.  No choice of $t_{shift}$  
can collapse such aging curves, and the scaling that is 
successful at $T>T_{stop}$ is not valid at low temperature.    Consequently, the behavior 
of the liquid at $T < T_{stop}$ should not be considered rejuvenation from the aging
at $T_{stop}$ in any sense, despite the susceptibility
ratio $\epsilon''_{stop}$/$\epsilon''_{ref}$
approaching one at low temperature.  The failure of the scaling through $t_{shift}$
indicates that the system at low temperature is in an out-of-equilibrium state unlike any 
that it would have entered without the cooling stop, which is contrary to the 
notion of rejuvenation.  Also, if one 
were to impose a $t_{shift}$ to match $\epsilon''$(0.79 Hz, $t_{a}=0$)
following a cooling stop with the corresponding value without a stop, 
one would require $t_{shift}>>t_{stop}$.  

\begin{figure}
    \centering\includegraphics[scale=1.0]{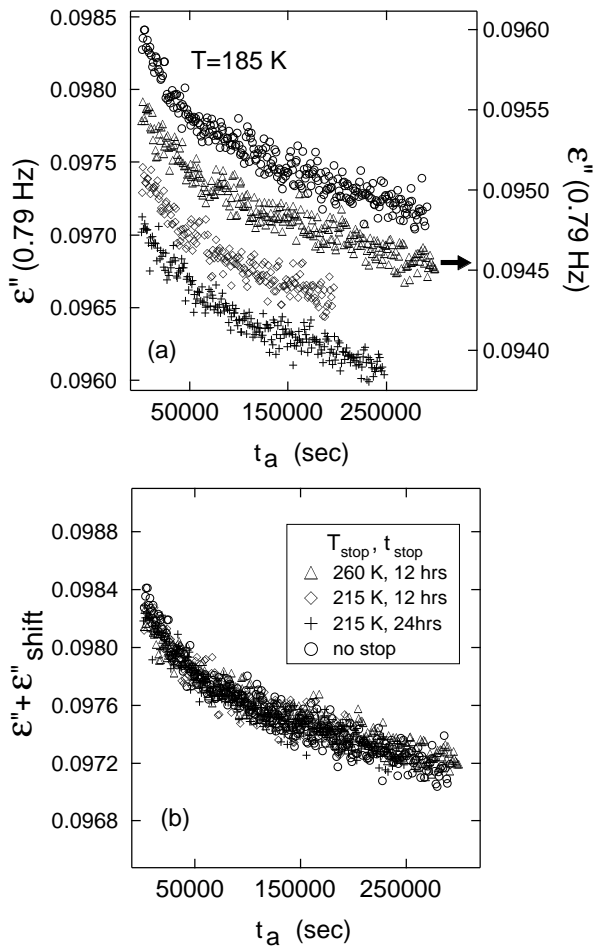}
\caption{\label{fig4} (a) $\epsilon$''(0.79 Hz) versus time, $t_{a}$, following an
interruption in reheating at 185 K during memory experiments that 
include a cooling stop at 
$T_{stop}=$260 K for $t_{stop}=$ 12 hours (triangles), 
$T_{stop}=$215 K for $t_{stop}=$ 12 hours (diamonds),
$T_{stop}=$215 K for $t_{stop}=$ 24 hours (crosses),
or no cooling stop (circles).  (b)  $\epsilon$''(0.79 Hz,$t_{a}$)
for these measurements shifted by additive constants, 
$\epsilon$''$_{shift}$, to collapse the results onto a single curve. }
\end{figure}

While the scaling through $t_{shift}$ fails for $T<T_{stop}$, the low
temperature aging curves, $\epsilon''$(0.79 Hz, $t_{a}$), 
can be superposed by introducing additive constants, $\epsilon''_{shift}$, 
that shift the curves vertically, as shown in Fig.~4(b)~\cite{multiply}.
This scaling through $\epsilon''_{shift}$ at $T < 
T_{stop}$ indicates that the aging rate, $d\epsilon''/dt_{a}$, is independent of 
the stopping.  Recent simulation studies have shown that 
aging in structural glasses occurs through a process of hopping among local minima with 
progressively lower energy within the potential energy landscape of the 
liquid~\cite{angelani,kob,donati}.  
Within such a picture, the failure of the scaling through $t_{shift}$ 
at $T<T_{stop}$ indicates that the aging at $T_{stop}$ directs the 
system to a location on the landscape after cooling to low temperature 
that it would never have visited had it been cooled directly to the low 
temperature and allowed to age (in contrast to when the 
temperature is raised back to $T>T_{stop}$).  However, the structure 
of the energy landscape must be such that the subsequent aging at low 
temperature is independent of the system's location on the landscape 
and dependent only on the system's temperature. 

In conclusion, by characterizing 
the aging of the molecular glass sorbitol following interruptions of
memory experiments, we have illuminated the out-of-equilibrium state of 
the system as it displays memory.  We find that the nature of the 
state depends crucially on whether the glass is 
at a temperature greater or less than $T_{stop}$.  This distinction, 
exemplified by the success of different scaling procedures in Figs.~3(b) and 
4(b), points to an interesting evolution of the system.  
Specifically, when the glass is reheated following a 
cooling that includes a stop, it goes from a state at low 
temperature unlike any that it would reach without the stop to a state at high 
temperature that has only an increased effective age due to the stop.
%While such an evolution suggests a picture of glassy sorbitol in 
%terms of an energy landscape with hierarchical structure, the 
%success of the activated scaling at high temperature also indicates that 
%these observations are fundamentally different from those
%expected for spin glasses.  
%Similar experiments on other structural 
%glasses would help clarify the generality of this 
%difference as well as test the speculative association between the dynamics 
%of the beta relaxation and aging.  
Further studies, particularly those that 
focus on the crossover between these high and low temperature scaling behaviors, 
would help clarify the significance of these findings.  While a smooth 
crossover between the two behaviors seems likely, its details and its 
position in temperature could be informative.  In addition, we note that 
for sorbitol $\epsilon$''(0.79 Hz) at the chosen values of $T_{stop}$ has a larger 
contribution from the beta 
relaxation at $T<T_{stop}$ than it does at $T>T_{stop}$.  While the strong 
similarity between memory curves for sorbitol and other glassy 
materials suggests that the beta 
relaxation is not playing a 
significant role, measurements on other molecular liquids without 
such a pronounced beta peak would help clarify its impact 
on the observed behavior.  Finally, we note that
simulations of structural glasses, which have recently 
placed aging in the context of energy 
landscape features, could further illuminate the significance 
of the evolution in scaling that occurs between $T<T_{stop}$ and $T>T_{stop}$.   
In particular, simulation studies that include more 
complicated thermal histories for the glass, such as those imposed 
experimentally in this work, could make concrete the features of the
energy landscape suggested by these measurements.

%\acknowledgments
We thank J.-P. Bouchaud for 
helpful discussions.  This work was supported by the National Science Foundation
under CAREER Award No.~DMR-0134377.


\begin{references}

\bibitem{struik}
STRUIK L. C. E., {\it Physical Aging in Amorphous Polymers and Other 
Materials} (Elsevier, Amsterdam, 1978).

\bibitem{matsuoka}
MATSUOKA S., {\it Relaxation Phenomena in Polymers} (Hanser, New 
York, 1992).

\bibitem{vincent-review}
VINCENT E., HAMMANN J., OCIO M., BOUCHAUD J.-P. and CUGLIANDOLO L. F.,
in {\it Complex Behavior of Glassy Systems}, 
edited by M. RUB\'{I} and C. PEREZ-VICENTE (Springer-Verlag, New York) 
1997.

\bibitem{nordblad-review}
NORDBLAD P., 
in {Dynamical properties of unconventional magnetic systems},
edited by A. T. SKJELTORP and D. SHERRINGTON (Kluwer, Boston) 1998.

\bibitem{leheny}
LEHENY R. L. and NAGEL S. R.,
{\it Phys. Rev. B}, {\bf 57} (2001) 5154.

\bibitem{cipelletti}  
CIPELLETTI L., MANLEY S., BALL R. C. and WEITZ D. A., 
{\it Phys. Rev. Lett.}, {\bf 84} (2000) 2275.

\bibitem{harden}
KNAEBEL A., BELLOUR M., MUNCH J. P., VIASNOFF V., LEQUEUX F. and 
HARDEN J. L.,
{\it Europhys. Lett.}, {\bf 52} (2000) 73.

\bibitem{ferroelectric-aging1}
ALBERICI F., DOUSSINEAU P. and LEVELUT A., 
{\it J. Phys. I France}, {\bf 7} (1997) 329.

\bibitem{ferroelectric-aging2}
COLLA E. V., CHAO L. K., WEISSMAN M. B. and VIEHLAND D. D.,
{\it Phys. Rev. Lett.}, {\bf 85} (2000) 3033.

\bibitem{bouchaud}
BOUCHAUD J.-P.,
in {\it Soft and Fragile Matter: Nonequilibrium Dynamics, Metastability and Flow}, 
edited by M. E. CATES and M. R. EVANS, (IOP, Bristol) 2000.

\bibitem{cugliandolo}
CUGLIANDOLO L. F., KURCHAN J. and PELITI L., 
{\it Phys. Rev. E}, {\bf 55} (1997) 3898. 

\bibitem{grigera}
GRIGERA T. S. and ISRAELOFF N. E., 
{\it Phys. Rev. Lett.}, {\bf 83} (1999) 5038. 

\bibitem{jonason}
JONASON K., VINCENT E., HAMMANN J., BOUCHAUD J.P. and NORDBLAD P., 
{\it Phys. Rev. Lett.}, {\bf 81} (1998) 3243.

\bibitem{bellon}
BELLON L., CILIBERTO S., LAROCHE C., 
{\it Europhys. Lett.}, {\bf 51} (2000) 551.

\bibitem{doussineau}
DOUSSINEAU P., De LACERDA-AR\^{O}SO T., and LEVELUT A., 
{\it Europhys. Lett.}, {\bf 46} (1999) 401.

\bibitem{beta}
YARDIMCI H. and LEHENY R. L.,
in preparation.

\bibitem{kovacs}
KOVACS A. J.,
{\it J. Polym. Sci.}, {\bf 30} (1958) 131.

\bibitem{refregier}
REFREGIER P., VINCENT E., HAMMANN J. and OCIO M.,
{J. Physique}, {\bf 48} (1987) 1533.

\bibitem{bellonEPJB}
BELLON L., CILIBERTO S., LAROCHE C., 
{Eur. Phys. J. B}, {\bf 25} (2002) 223.

\bibitem{multiply}
Given the relative values of $\epsilon''$(0.79 Hz, $t_{a}$) and the 
similar shapes of the curves for different $T_{stop}$ and $t_{stop}$
in Fig.~4(a), we find that the aging curves at $T<T_{stop}$ can be scaled equally 
well through application of a multiplicitive factor as they are 
through an additive factor.

\bibitem{angelani}
ANGELANI L., Di LEONARDO R., PARISI G. and RUOCCO, G.
{\it Phys. Rev. Lett.}, {\bf 87} (2001) 055502.

\bibitem{kob}
KOB W., SCIORTINO F. and TARTAGLIA P.,
{\it Europhys. Lett.}, {\bf 49} (2000) 590.

\bibitem{donati}
DONATI C., SCIORTINO F. and TARTAGLIA P.,
{\it Phys. Rev. Lett.}, {\bf 85} (2000) 1464.

\end{references}
\end{document}